\documentclass[twocolumn,aps,prl,draft,showpacs]{revtex4}
\usepackage{graphicx}
\usepackage{bm}
\begin{document}
\def\be{\begin{equation}}
\def\ee{\end{equation}}
\def\bea{\begin{eqnarray}}
\def\eea{\end{eqnarray}}
\def\E{{\rm e}}
\def\bearst{\begin{eqnarray*}}
\def\eearst{\end{eqnarray*}}
\def\peleven{\parbox{11cm}}
\def\peffec{\peight{\bearst\eearst}\hfill\peleven}
\def\pspace{\peight{\bearst\eearst}\hfill}
\def\ptwelve{\parbox{12cm}}
\def\peight{\parbox{8mm}}

\title{Lagrangian tracers on a surface flow: the role of time correlations}
 \author{Guido~Boffetta$^{1}$, Jahanshah~Davoudi$^{2}$, 
Bruno~Eckhardt$^{2}$ and J\"org Schumacher$^{2}$}
\affiliation{$^{1}$ Dipartimento di Fisica Generale and INFM,
Universit\`a di Torino,
Via Pietro Giuria 1, I-10125 Torino, Italy}
\affiliation{ $^{2}$ 
Fachbereich Physik, Philipps-Universit\"at, Renthof 6,
D-35032 Marburg, Germany}            

\begin{abstract}
Finite time correlations of the velocity in a surface flow 
are found to be important for the formation of clusters of Lagrangian tracers. 
The degree of clustering characterized  
by the Lyapunov spectrum of the flow is numerically shown to be in qualitative
agreement with the predictions for the white-in-time compressible
Kraichnan flow, but to deviate quantitatively.
For intermediate values of compressibility the clustering is 
surprisingly weakened by time correlations.



\end{abstract}
\pacs{47.27.-i, 47.27.Ak, 05.40.-a}
\maketitle
Inhomogeneous distribution of particles advected
in a turbulent flow is a generic consequence
of compressibility. This can be obtained in two
situations. The first possibility is that 
the advecting flow is compressible itself and that
the particles follow the streamlines \cite{Sommerer93}. The other
possibility is that the particles do not follow
the streamlines because of inertia \cite{Sundaram97,Klyatskin97,BFF01,Benczik02,Lvov2002,Mehlig03,Bec03} or 
lift \cite{mazzitelli03},
and that the effective velocity field is compressible.
Such situations are relevant for the formation
of clouds \cite{Shaw03} or for the advection of bubbles in turbulent flows, 
e.g.  for breaking waves on
the ocean surface \cite{Magnaudet00,Deane02}. 
We focus here on the first possibility.

While the dominant tendency of incompressible flows is to separate particle
trajectories, a compressible component is responsible for particle trapping in
contracting regions for long times.  The Eulerian compressibility of a flow is
measured by the dimensionless ratio ${\cal C} = {\langle \left( \partial_i u_i
\right)^2 \rangle /\langle \left( \partial_i u_j \right)^2 \rangle }$. It takes
values between 0 (incompressible flow) and 1 (potential flow).
While there can be
no clustering without compressible effects,
the compressibility ratio ${\cal C}$ is insufficient
to determine the final distribution completely \cite{FGV01}:
the behavior depends also on the spatial
roughness, the dimensionality, and, as we
will demonstrate here, on the time correlations
in the flow.

A convenient characterization of the final
distribution uses the Lagrangian Lyapunov spectrum.
Dynamical systems theory shows that the 
asymptotic clusters are
smooth along the unstable directions of positive
Lyapunov exponents and fractal along the stable
directions.
In $d$ dimensions, the sum of Lyapunov exponents 
$\sum_{i=1}^d \lambda_i$ vanishes in the 
incompressible case and is negative for a compressible
flow. A measure for the distribution
of the final clusters is given by the Lyapunov dimension \cite{ott},
\begin{equation}
D_{L}=K+{\sum_{i=1}^{K} \lambda_i \over |\lambda_{K+1}|}\,,
\label{eq:2}
\end{equation}
where K is the maximal integer such that $\sum_{i=1}^K \lambda_i \ge 0$.
Obviously, for an incompressible flow $K=d$ and
also $D_L=d$: the particle distribution fills the
entire volume. As compressibility increases,
the sum of Lyapunov exponents will
become negative and $K$ will drop below $d$.
If the largest Lyapunov exponent $\lambda_1$ becomes
negative, the final cluster will not have
any smooth directions anymore and the particles
will cluster in a point-like fractal. One
hence distinguishes a regime of strong
compressibility $\lambda_1<0$ with $D_{L}=0$, and one of
weak compressibility $\lambda_1>0$.

There are no general results on the spectrum of Lagrangian Lyapunov exponents in
turbulent flows.  For the case of incompressible, isotropic, three dimensional
turbulence the numerical observation is that $\lambda_2 \sim 1/4 \lambda_1$
\cite{GP90}. In the limit of a compressible Kraichnan flow, which is 
a synthetic, white-in-time, and Gaussian distributed, the spectrum is given by
$\lambda_j=C_1 \left[d(d-2j +1) - 2{\cal C}(d+(d-2)j)\right]$ where 
$j=1,...,d$, $C_1$ is an
inverse time proportional to the Lagrangian strain and the
resulting Lyapunov dimension is a decreasing function of ${\cal C}$
\cite{FGV01}.  Moreover, when ${\cal C} \ge {\cal C}_s = d/4$ 
all the Lyapunov exponents become negative and 
one has the strong compressible regime.

The trapping effects are believed to be enhanced in spatially {\it rough}
Kraichnan flows \cite{FGV01}.  Although a recent attempt has been devoted to
re-introduce the finite time correlation for the synthetic rough Kraichnan flows
\cite{Chaves03} a general theoretical framework studying the effects of the time
correlation is still lacking.

\begin{figure}[htb]
\centerline{\includegraphics[angle=0,scale=0.4,draft=false]{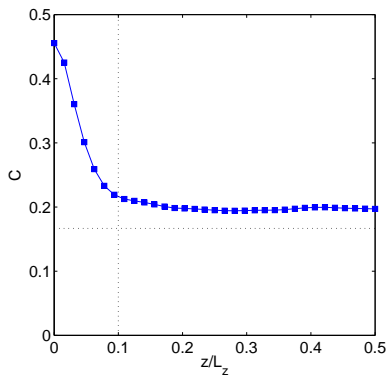}}
\narrowtext
\caption{Compressibility ${\cal C}$ as a function of depth $z/L_z$ ranging
from the free surface ($z/L_z=0$) to the middle of the tank ($z/L_z=0.5$). 
The figure illustrates also that for $z/L\ge 0.1$ the volume turbulence 
gets close to the homogeneous isotropic case which is indicated by the
vertical dotted line. The horizontal dotted line indicates
the ${\cal C}$ value of 1/6 which would follow for perfect isotropy in the 
bulk.}
\label{fig1}
\end{figure}


Here, we demonstrate numerically the crucial role of finite time 
correlations in the Lagrangian statistics of particles transported by the
Navier-Stokes flow which is established in a free-slip surface. 
As a consequence of boundary conditions the two-dimensional flow
displays an effective compressibility.
By decomposing the surface velocity field in its irrotational and
potential components, we change the effective value of compressibility
of the surface flow.
This enables us to study the interplay of the two components, 
time correlations and compressibility, within one system.

We want to remark that our investigation does not encounter the effects of 
fluid density variations which appear in a compressible turbulent flow. 

The tracers are advected by the free surface flow on top of three-dimensional
($3d$) incompressible Navier-Stokes turbulence at $Re_{\lambda} \simeq 145$. 
The volume with an aspect ratio of $2\pi:2\pi:1$ is resolved by 
$256\times 256\times 65$ grid points. The full $3d$
equations are integrated by a standard pseudo-spectral method \cite{PRE}. 
The lateral boundary conditions
for $x$ and $y$ are periodic and in the vertical direction we apply free-slip
boundary conditions, $u_z=0$ and $\partial_z u_x=\partial_z u_y=0$.  The
Lagrangian tracers are advected by the surface flow spanned by the two
components $u_x(x,y,z=0,t)$ and $u_y(x,y,z=0,t)$ which is compressible since
$\partial_x u_x + \partial_y u_y = - \partial_z u_z \neq 0$ \cite{PRE}. Tracer velocity between the grid mesh is 
calculated by bi-linear interpolation. We
store 18,000 configurations of the $2d$ surface velocity field at the same resolution
in time that is used for the time advancement of the full $3d$ Navier-Stokes
equations.  On the basis of these configurations, Lagrangian trajectories and the
Lyapunov spectrum $(\lambda_1,\lambda_2)$ are computed.

Isotropy can be used to show that the compressibility ratio
for a two-dimensional flow $(u_x(x,y),u_y(x,y))$
can be rewritten as
\begin{equation}
{\cal C} = {\langle \left(\partial_x u_x + \partial_y u_y \right)^2 \rangle
\over
2 \langle \left(\partial_x u_x - \partial_y u_y \right)^2 \rangle } \,.
\label{eq:1}
\end{equation}
Numerical and experimental
investigations of the surface flow at Taylor Reynolds numbers $Re_\lambda \sim
10^2$ have shown that ${\cal C} \simeq 0.5$ \cite{walter,Schu02}.  By
(\ref{eq:1}), this is equivalent to a vanishing value of the cross correlation $\langle
\partial_x u_x \partial_y u_y\rangle$, as confirmed in numerical studies \cite{NJP}.  
Figure~\ref{fig1} shows the profile of
${\cal C}$ for all planes of the simulation box from
$z=0$ down to $z=L/2$.  It is interesting to note that the value of
compressibility measured in the bulk is very close to the value $1/6$ which is
expected for a two-dimensional cut of a three-dimensional homogeneous, isotropic
flow.  The relatively large value of the compressibility ratio on the free 
surface is a consequence of the free-slip boundary conditions.

\begin{figure}[htb]
\centerline{\hspace{-0.0cm}
\includegraphics[angle=0,scale=0.6,draft=false]{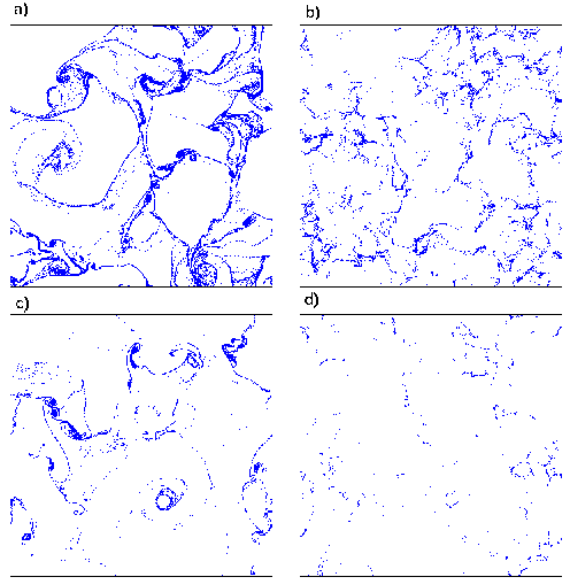}}
\narrowtext
\caption{Visualization of the particle clustering for a uniform initial
distribution of 36,000 tracers on the free surface $z=0$ with ${\cal C}=0.45$.
The panels ($a$) and ($c$) show the tracer distributions in time correlated
surface flow at $0.2 T$ and $3.1 T$.  The panels ($b$) and ($d$) correspond to
the tracer distributions in the time de-correlated flow at $0.2 T$ and $5.1 T$.
The time scale $T=L/\delta u(L)$ is the large scale eddy turnover time.}
\label{fig2}
\end{figure}

The evolution of an initially uniform particle distribution in the 
time-correlated surface flow with ${\cal C}=0.45$ is shown at two instants in 
Figs.~\ref{fig2}$a$ and \ref{fig2}$c$, respectively.
The pronounced formation of particle clusters was observed recently in
experiments \cite{walter} and in simulations \cite{Schu02}.

\begin{figure}[htb]
\centerline{\includegraphics[scale=0.7,draft=false]{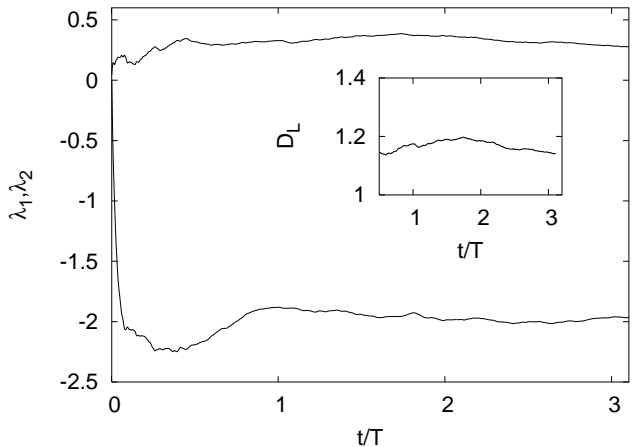}}
\caption{Convergence of Lyapunov exponents ($\lambda_1>0$ and 
$\lambda_2<0$) for the Lagrangian 
trajectories on the free surface computed for about three eddy
turnover times $T$. 
In the inset, the Lyapunov dimension
$D_L=1+\lambda_1/|\lambda_2|$ is shown.}
\label{fig3}
\end{figure}

Although having a compressibility close to the critical value of 
${\cal C}_2=1/2$ for the
two-dimensional Kraichnan flow we do not observe point-like structures;
particles accumulate on a network of narrow ridges with bigger empty voids in
between.  Figure~\ref{fig3} reports the results of the numerical computation of
the Lagrangian Lyapunov spectrum.  The very
existence of the ridge structures pertains to the existence of a finite positive
Lyapunov exponent $\lambda_1\sim 0.3$.  This is in line with the computed
Lyapunov dimension, $D_L \simeq 1.15$ which is plotted in the inset of
Fig.~\ref{fig3}. 

We now compare the Lyapunov dimensions of the time-correlated flow with the 
the corresponding de-correlated flow. 
That would quantify the effects of time correlation effects on clustering 
properties.
In order to do so, we generate the decorrelated velocity
field from the flow at hand by re-shuffling the temporal
sequence of surface flow snapshots. 
Thanks to the periodic boundary conditions, each snapshot is
further de-correlated by phase randomization in space.

\begin{figure}[htb]
\centerline{\includegraphics[scale=0.7,draft=false]{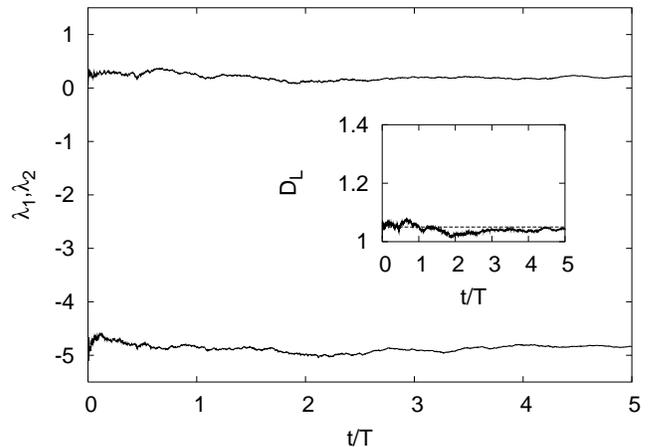}}
\caption{Convergence of Lyapunov exponents for the Lagrangian 
trajectories advected by the random-time, random-phase
scrambled velocity field. The computation is done for $5 T$.
In the inset, we show the Lyapunov dimension $D_L$ together with
the theoretical prediction $D_L = 1.05$ (dashed line).}
\label{fig4}
\end{figure}

Figures~\ref{fig2}$b$ and~\ref{fig2}$d$ show the evolution of the density of
tracers in the re-shuffled time de-correlated flow.  We find the
clusters concentrated on more point-like sets and the ridges are depleted
evidently.  Therefore one expects that the Lyapunov dimension is {\it
decreased} when the temporal correlation of the velocity is diminished.
Lagrangian trajectories in the randomized flow are more erratic:  in order to
achieve statistical convergence, trajectories are now integrated for about $5$
large scale eddy turnover times.  The computation of Lyapunov spectrum and
dimension for the randomized flow is shown in Figure~\ref{fig4}.  The fact that
both of the Lyapunov exponents are changed with respect to Fig.~\ref{fig3} is
not relevant, as the characteristic Lagrangian time is changed by time
decorrelation procedure.  The important point is that the Lyapunov dimension is
now $D_L \simeq 1.05$  which is {\it less} than the Lyapunov dimension in 
the time correlated flow.
This is quite surprising as one intuitively expects that the particle 
trapping is
enhanced when moving in a time-correlated flow because the local flow 
will persist for some time.  

In the case of two-dimensional compressible Kraichnan ensemble the pairing
symmetry in the Lyapunov spectrum ($\lambda,-\lambda$) is broken and the
Lyapunov dimension is then simply given by
\begin{equation}
D_L = {2 \over 1 + 2 {\cal C}}\,.
\label{eq:3}
\end{equation}
 According to
(\ref{eq:3}), the transition to the strong compressibility regime 
takes place at ${\cal C}_s=1/2$, very close
to the numerically observed compressibility at the surface (see
Fig.~\ref{fig1}).  The prediction of (\ref{eq:3}) for ${\cal C}\simeq 0.45$
renders $D_L \sim 1.06$ which is in very good agreement with the result of randomized surface flow.
The compressibility, seen along Lagrangian trajectories \cite{BFF01}, is
different from ${\cal C}$ as given by (\ref{eq:1}).  However explicit
theoretical results inferring to the effects of velocity time correlations on
effective Lagrangian compressibility or Lyapunov spectrum are still lacking
although first attempts have been made \cite{Chaves03}.  Moreover, it is not
clear that the theoretical description of the time correlated flows will enjoy
the same level of generality.

In order to gain more insight on the role of the time correlations we repeated
the Lagrangian analysis for different degrees of compressibility.
To this end, we have decomposed the $2d$ velocity snapshots into the stream 
function $\psi(x,y)$ and the
potential function $\phi(x,y)$.  A new velocity field is
reconstructed then to
\be
\tilde{\bf u}(x,y) = \sqrt{2} \left({\bf \nabla}^{T} \psi(x,y) \cos \gamma +
{\bf \nabla} \phi(x,y) \sin \gamma \right)
\label{eq:4}
\ee
where ${\bf \nabla}^T=(\partial_y,-\partial_x)$.  The compressibility 
ratio ${\cal C}$
(cf.~(\ref{eq:1})) for the decomposed flows is a monotonic function of the
parameter $\gamma \in [0,\pi/2]$ and for $\gamma=\pi/4$ the original
surface flow is reproduced.  $\gamma=0$ corresponds to the incompressible case and
$\gamma=\pi/2$ gives a purely potential flow; so we are able to cover the whole
range of compressibility degrees.  Of course, dimensional quantities such as the
characteristic Lagrangian time of the velocity field may depend on the value of
$\gamma$.  On the other hand, dimensionless quantities, such as $D_L$ are
independent on a global rescaling of the velocity and thus can be expected to
depend mainly on the degree of compressibility.

\begin{figure}[htb]
\centerline{\includegraphics[scale=0.65,draft=false]{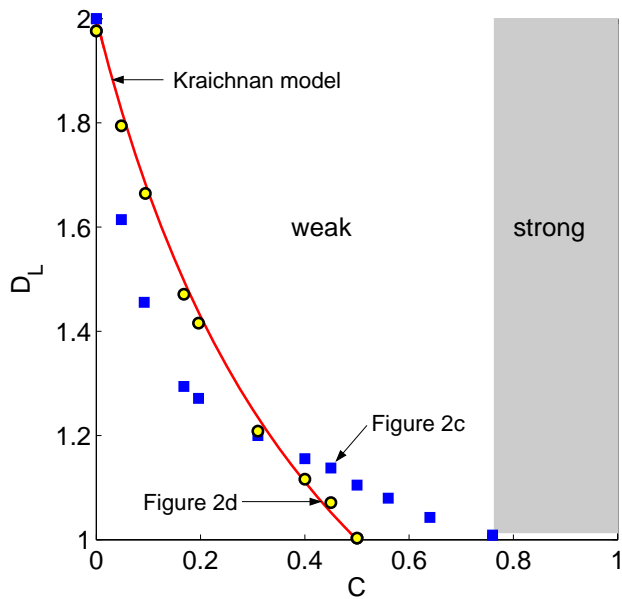}}
\caption{Lyapunov dimension $D_L$ computed for different values of the
compressibility ${\cal C}$ in accordance with (\ref{eq:4}).  Data points for
time-correlated flow are indicated by squares.  The open circles are for the
de-correlated flow.  The line indicates the behavior for the compressible
Kraichnan flow (cf.~(3)).  The gray shading in the background illustrates the
border between weak and strong compressible regime for the time-correlated flow.
We have also indicated the data points in the plane that correspond with the
tracer distributions as given in Fig.~2.} 
\label{fig5}
\end{figure}

We have performed two sets of numerical simulations of Lagrangian trajectories
for different values of $\gamma$.  For each value, we have computed the mean
compressibility and the Lyapunov dimension, both for the time-correlated and
randomized flows.  The result is summarized in Fig.~\ref{fig5} where for
comparison we plot the relation (\ref{eq:3}) for the $\delta$-correlated
Kraichnan case.

It is remarkable that the qualitative effect of time correlations on tracers
distribution depends on the level of compressibility.  For small values ${\cal
C} \leq 0.3$, time correlations enhance the effects of compressibility, i.e.
$D_L$ smaller than the predicted value by (\ref{eq:3}).  The situation for
larger values of ${\cal C}$ is more surprising as the Lyapunov dimension in
correlated flow is larger than the value given by (\ref{eq:3}), i.e.  the
effects of compressibility are depleted with respect to the $\delta$-correlated
case.  The later situation is contrary to intuition according to which finite
time correlations amplify the effects of compressibility.  Consequently the
transition point to the strong compressible regime, where $\lambda_1 \le 0$,
moves up to the ${\cal C} \simeq 0.75$ larger than the prediction of
(\ref{eq:3}) (cf.  gray region in Fig.~\ref{fig5}).  This is again
the point where the ratio $\frac{\lambda_1}{\lambda_2}$ gets zero for the
correlated flow.  

In summary, we have studied the effects of time correlations in the clustering of Lagrangian 
tracers advected on a free surface flow by direct numerical simulation of NS equations. 
Appealing to a numerical
decomposition method we generate $2d$ flows with variable degrees of
compressibilities.  The Lyapunov dimension decreases with the compressibility
degree.  Our simulations indicate that the effect of time correlation can go
both ways.  At low values of compressibility the Lyapunov dimension of the
tracer distribution is smaller than the dimension in the time de-correlated
flows.  Yet as the compressibility gets higher this regime crosses over to a
more interesting one where the time correlations increase the Lyapunov
dimension.  The point of phase transition to strong compressible regime is at a
higher value, ${\cal C}\simeq 0.75$. The present paper is a first attempt to relate aspects  
of the solvable compressible Kraichnan model to the turbulent Navier-Stokes compressible flows. This
suggests further investigations on the interrelated roles of the time correlations and the compressibility on the 
dispersion properties. 

This work was supported by the Deutsche Forschungsgemeinschaft.
The computations were carried out at the John von Neumann Institute 
for Computing at the Forschungszentrum J\"ulich and we are grateful for
support. ~G.~B. acknowledges hospitality of the Philipps Universit\"at 
Marburg.

\end{document}